\documentclass[12pt]{iopart}
\usepackage{graphics}

\begin{document}

\title[The hyperbolic, the arithmetic and the quantum phase]
{The hyperbolic, the arithmetic \\and the quantum phase}


\author{Michel Planat\dag\ \footnote[3]{To whom correspondence should be addressed
(planat@lpmo.edu)} and Haret Rosu\ddag  }

\address{\dag\ Laboratoire de Physique et M\'{e}trologie des Oscillateurs du
CNRS,\\ 32 Avenue de l'observatoire, 25044 Besan\c{c}on Cedex,
France\\}

\address{\ddag\ Potosinian Institute of Scientific and Technological Research\\
Apdo Postal 3-74, Tangamanga, San Luis Potosi, SLP, Mexico\\}

\begin{abstract}
 We develop a new approach of the quantum phase in an
Hilbert space of $\rm{fi}$nite dimension which is based on the
relation between the physical concept of phase locking and
mathematical concepts such as cyclotomy and the Ramanujan sums. As
a result, phase variability looks quite similar to its classical
counterpart, having peaks at dimensions equal to a power of a
prime number. Squeezing of the phase noise is allowed for
speci$\rm{fi}$c quantum states. The concept of phase entanglement
for Kloosterman pairs of phase-locked states is introduced.

\end{abstract}


\maketitle

\section{Introduction}
Time and phase are not well de$\rm{fi}$ned concepts at the quantum
level. The present work belongs to a longstanding effort to model
phase noise and phase-locking effects that are found in highly
stable oscillators. It was unambiguously demonstrated that the
observed variability, (i.e., the 1/f frequency noise of such
oscillators) is related to the $\rm{fi}$nite dynamics of states
during the measurement process and to the precise $\rm{fi}$ltering
rules that involve continued fraction expansions, prime number
decomposition, and hyperbolic geometry
\cite{Planat1}-\cite{Planat3}.

We add here a quantum counterpart to these effects by studying the
 notions of quantum phase-locking and quantum phase
entanglement. The problem of de$\rm{fi}$ning quantum phase
operators was initiated by Dirac in 1927 \cite{Dirac}. For
excellent reviews, see \cite{Reviews}. We use the Pegg and Barnett
quantum phase formalism \cite{Pegg} where the calculations are
performed in an Hilbert space $H_q$ of $\rm{fi}$nite dimension q.
The phase states are de$\rm{fi}$ned as superpositions of number
states from the so-called quantum Fourier transform (or QFT)
\begin{equation}
|\theta _p\rangle=q^{-1/2}\sum_{n=0}^{q-1}\exp(\frac{2i\pi p n
}{q})|n\rangle~. \label{eq1}
\end{equation}
in which $i^2=-1$. The states $|\theta_p\rangle$ form an
orthonormal set and in addition the projector over the subspace of
phase states is $\sum_{p=0}^{q-1}|\theta_p\rangle
\langle\theta_p|=1_q$ where $1_q$ is the identity operator in
$H_q$. The inverse quantum Fourier transform follows as
$|n\rangle=q^{-1/2}\sum_{p=0}^{q-1}\exp(-\frac{2i\pi p n
}{q})|\theta _p\rangle$. As the set of number states $|n\rangle$,
the set of phase states $|\theta_p\rangle$ is a complete set
spanning $H_q$. In addition the QFT operator is a $q$ by $q$
unitary matrix with matrix elements
$\kappa_{pn}^{(q)}=\frac{1}{\sqrt {q}}\exp(2i\pi \frac{pn}{q})$.

From now we emphasize phase states $|\theta' _p\rangle$ satisfying
phase-locking properties. The quantum phase-locking operator is
defined in (\ref{eq4}). We $\rm{fi}$rst impose the coprimality
condition
\begin{equation}
(p,q)=1,
\end{equation}
where $(p,q)$ is the greatest common divisor of $p$ and $q$.
Differently from the phase states (\ref{eq1}), the $|\theta'
_p\rangle$ form an orthonormal base of a Hilbert space whose
dimension is lower, and equals the number of irreducible fractions
$p/q$, which is given by the Euler totient function $\phi(q)$.
These states were studied in our recent publication
\cite{PLA03}\footnote{ Some errors or misunderstandings are
present in that earlier report. The summation in (3),(5),(7) and
(9) should  be (this is implicit) from $0$ to $\phi(q)$. The
expectation value $\langle \theta_q^{\rm{lock}}\rangle$ in (8)
should be squared. There are also slight changes in the plots.}.

Guided by the analogy with the classical situation \cite{Planat2},
we call these irreducible states the phase-locked quantum states.
They generate a cyclotomic lattice $L$ \cite{Craig} with generator
matrix $M$ of matrix elements $\kappa_{pn}^{'(q)}$, (p,q)=1 and of
size $\phi(q)$. The corresponding Gram matrix $H=M^{\dag}M$ shows
matrix elements $h_{n,l}^{(q)}=c_q(n-l)$ which are Ramanujan sums
\begin{equation}
c_q(n)=\sum_p \exp(2i\pi \frac{p}{q}
n)=\frac{\mu(q_1)\phi(q)}{\phi(q_1)},~~\rm{with}~q_1=q/(q,n).
\label{eq2}
\end{equation}
where the index $p$ means summation from $0$ to $q-1$, and
$(p,q)=1$. Ramanujan sums are thus de$\rm{fi}$ned as the sums over
the primitive characters $\exp(2i\pi \frac{pn}{q})$, (p,q)=1, of
the group $Z_q=Z/qZ$. In the equation above $\mu(q)$ is the
M\"obius function, which is $0$ if the prime number decomposition
of $q$ contains a square, $1$ if $q=1$, and $(-1)^k$ if $q$ is the
product of $k$ distinct primes \cite{Hardy}. Ramanujan sums are
relative integers which are quasi-periodic versus $n$ with
quasi-period $\phi(q)$ and aperiodic versus $q$ with a type of
variability imposed by the M\"obius function. Ramanujan sums were
introduced by Ramanujan in the context of Goldbach conjecture
\cite{Hardy}.

They are also useful in the context of signal processing as an
arithmetical alternative to the discrete Fourier transform
\cite{Planat4}. In the discrete Fourier transform the signal
processing is performed by using all roots of unity of the form
$\exp(2i\pi p/q)$ with $p$ from $1$ to $q$ and taking their nth
powers $e_p(n)$ as basis function. We generalized the classical
Fourier analysis by using Ramanujan sums $c_q(n)$ as in
(\ref{eq2}) instead of $e_p(n)$. This type of signal processing is
more appropriate for arithmetical functions than is the ordinary
discrete Fourier transform, while still preserving the metric and
orthogonal properties of the latter. Notable results relating
arithmetical functions to each other can be obtained using
Ramanujan sums expansion while the discrete Fourier transform
would show instead the low frequency tails in the power spectrum.

In this paper we are also interested in pairs of phase-locked
states which satisfy the two conditions
\begin{equation}
(p,q)=1~~and~~p\bar{p}=-1 (mod~q).
 \label{phaselocked}
\end{equation}
Whenever it exists $\bar{p}$ is uniquely de$\rm{fi}$ned from minus
the inverse of $p$ modulo $q$. Geometrically the two fractions
$p/q$ and $\bar{p}/q$ are the ones selected from the partition of
the half plane by Ford circles. Ford circles are de$\rm{fi}$ned as
the set of the images of the horizontal line z=x+i, x real, under
all modular transformations in the group of $2\times 2$ matrices
$SL(2,Z)$ \cite{Planat3}. Ford circles are tangent to the real
axis at a Farey fraction $p/q$, and are tangent to each other.
They have been introduced by Rademacher as an optimum integration
path to compute the number of partitions by means of the so-called
Ramanujan's circle method. In that method two circles of indices
$p/q$ and $\bar{p}/q$, of the same radius $\frac{1}{2q^2}$ are
dual to each other on the integration path [see also Part 2 and
Fig. 3].

\section{Hints to the hyperbolic geometry of phase noise}
\subsection{The  phase-locked loop, low pass filtering and $1/f$ noise}
A newly discovered clue for $1/f$ noise was found from the concept
of a phase locked loop (or PLL) \cite{Planat2}. In essence two
interacting oscillators, whatever their origin, attempt to
cooperate by locking their frequency and their phase. They can do
it by exchanging continuously tiny amounts of energy, so that both
the coupling coeff\mbox{}icient and the beat frequency should
f\mbox{}luctuate in time around their average value. Correlations
between amplitude and frequency noises were observed
\cite{Yamoto}.

One can get a good level of understanding of phase locking by
considering the case of quartz crystal oscillators used in high
frequency synthesizers, ultrastable clocks and communication
engineering (e.g., mobile phones). The PLL used in a FM radio
receiver is a genuine generator of $1/f$ noise. Close to phase
locking the level of $1/f$ noise scales approximately as
$\tilde{\sigma}^2$, where $\tilde{\sigma}=\sigma
K/\tilde{\omega_B}$ is the ratio between the open loop gain $K$
and the beat frequency $\tilde{\omega_B}$ times a
coeff\mbox{}icient $\sigma$ whose origin has to be explained. The
relation above is explained from a simple non linear model of the
PLL known as Adler's equation
\begin{equation}
\dot{\theta}(t)+K H(P)\sin \theta(t)=\omega_B, \label{equation1}
\end{equation}
where at this stage $H(P)=1$, $\omega_B=\omega(t)-\omega_0$ is the
angular frequency shift between the two quartz oscillators at the
input of the non linear device (a Schottky diode double balanced
mixer), and $\theta(t)$ is the phase shift of the two oscillators
versus time t. Solving (\ref {equation1}) and differentiating one
gets the observed noise level $\tilde{\sigma}$ versus the bare one
$\sigma=\delta \omega_B/\tilde{\omega_B}$. Thus the model doesn't
explain the existence of $1/f$ noise but correctly predicts its
dependence on the physical parameters of the loop \cite{Planat2}.

Besides one can get detailed knowledge of harmonic conversions in
the PLL by accounting for the transfer function $H(P)$, where
$P=\frac{d}{dt}$ is the Laplace operator. If $H(P)$ is a low pass
f\mbox{}iltering function with cut-off frequency $f_c$, the
frequency at the output of the mixer + f\mbox{}ilter stage is such
that
\begin{equation}
\mu=f_B(t)/f_0=q_i|\nu-p_i/q_i| \le f_c/f_0,~~
\mbox{$p_i$~and~$q_i$~integers}. \label{equation2}
\end{equation}
The beat frequency $f_B(t)$ results from the continued fraction
expansion of the input frequency ratio
\begin{equation}
\nu=f(t)/f_0=[a_0;a_1,a_2,\ldots a_i,a,\ldots ], \\
\label{equation3}
\end{equation}
where the brackets mean expansions
$a_0+1/(a_1+1/(a_2+1/\ldots+1/(a_i+1/(a+\ldots))))$. The
truncation at the integer part $a=[\frac{f_0}{f_c q_i}]$
def\mbox{}ines the edges of the basin for the resonance $p_i/q_i$;
they are located at $\nu_1=[a_0;a_1,a_2,\ldots,a_i, a]$ and
$\nu_2=[a_0;a_1,a_2,\ldots a_i-1,1,a]$ \cite{Planat1}. The two
expansions in $\nu_1$ and $\nu_2$, prior to the last
f\mbox{}iltering partial quotient $a$, are the two allowed ones
for a rational number. The convergents $p_i/q_i$ at level $i$ are
obtained using the matrix products
\begin{equation}
\left[\begin{array}{cc} a_0 & 1\\ 1 & 0 \end{array}\right]
\left[\begin{array}{cc} a_1 & 1\\ 1 & 0 \end{array}\right]\cdots
\left[\begin{array}{cc} a_i & 1\\ 1 & 0 \end{array}\right]
=\left[\begin{array}{cc} p_i&p_{i-1}\\ q_i&q_{i-1}
\end{array}\right]. \label{matrices}
\end{equation}
Using (\ref{matrices}), one can get the fractions $\nu_1$ and
$\nu_2$ as $\nu_1=\frac{p_a}{q_a}$ and
$\nu_2=\frac{p_i(2a+1)-p_a}{q_i(2a+1)-q_a}$, so that with the
relation relating convergents  $(p_i
q_{i-1}-p_{i-1}q_i)=(-1)^{i-1}$, the width of the basin of index
$i$ is
$|\nu_1-\nu_2|=\frac{2a+1}{q_a(q_a+(2a+1)q_i)}\simeq\frac{1}{q_a
q_i}$ whenever $a>1$.

 In previous publications of one of the authors a phenomenological model
 for $1/f$ noise in the PLL was proposed, based on an arithmetical function
 which is a logarithmic coding for prime numbers \cite{Planat1},\cite{Planat2}. If one accepts a
 coupling coeff\mbox{}icient evolving discontinuously versus the time $n$
 as $K=K_0\Lambda(n)$, with $\Lambda(n)$ the Mangoldt function
which is $\ln(p)$ if $n$ is the power of a prime number $p$ and
$0$ otherwise, then the average coupling coeff\mbox{}icient is
$K_0$ and there is an arithmetical f\mbox{}luctuation
$\epsilon(t)$
\begin{eqnarray}
&\psi(t)=\sum_{n=1}^t\Lambda(n)=t(1+\epsilon(t)),\nonumber\\
&t\epsilon(t)=-\ln(2\pi)-\frac{1}{2}\ln(1-t^{-2})-\sum_{\rho}\frac{t^{\rho}}{\rho}.
\label{Riemann}
\end{eqnarray}
The three terms at the right hand side of $t\epsilon(t)$ come from
the singularities of the Riemann zeta function $\zeta(s)$, that
are the pole at $s=1$, the trivial zeros at $s=-2l$, $l$ integer,
and the zeros on the critical line
$\Re(s)=\frac{1}{2}$\cite{Planat1}. Moreover the power spectral
density roughly shows a $1/f$ dependance versus the Fourier
frequency $f$. This is the proposed relation between Riemann zeros
(the still unproved Riemann hypothesis is that all zeros should
lie on the critical line) and $1/f$ noise.

We improved the model by replacing the Mangoldt function by its
modif\mbox{}ied form $b(n)=\Lambda(n)\phi(n)/n$, with $\phi(n)$
the Euler (totient) function \cite{Planat4}. This seemingly
insignif\mbox{}icant change was introduced by Hardy \cite{Hardy}
in the context of Ramanujan sums for the Goldbach conjecture and
resurrected by Gadiyar and Padma in their recent analysis of the
distribution of pairs of prime numbers \cite{Gadiyar}. Then by
defining the error term $\epsilon_B(t)$ from the cumulative
modified Mangoldt function
\begin{equation}
B(t)=\sum_{n=1}^t b(n)=t(1+\epsilon_B(t)), \label{functionB}
\end{equation}
its power spectral density $S_B(f)\simeq\frac{1}{f^{2\alpha}}$
exhibits a slope close to the Golden ratio
$\alpha\simeq(\sqrt5-1)/2\simeq0.618$ (see Fig. 1).

\begin{figure}[htbp]
\centering{\resizebox{10cm}{!}{\includegraphics{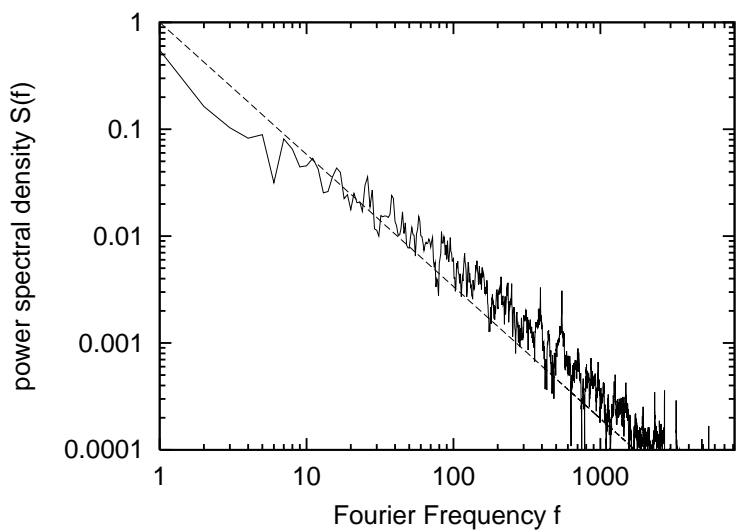}}}
\caption{Power spectral density of the error term in the
modif\mbox{}ied Mangoldt function b(n) in comparison to the power
law $1/f^{2\alpha}$, with the Golden ratio
$\alpha=(\sqrt{5}-1)/2$.}
\end{figure}
%

%
%
The modif\mbox{}ied Mangoldt function occurs in a natural way from
the logarithmic derivative of the following quotient
\begin{equation}
Z(s)=\frac{\zeta(s)}{\zeta(s+1)}=\sum_{n\ge
1}\frac{\phi(n)}{n^{s+1}}, \label{equation11}
\end{equation}
since $-\frac{Z'(s)}{Z(s)}=\sum_{n\ge 1}\frac{b(n)}{n^s}$. This
replaces the similar relation from the Riemann zeta function where
$-\frac{\zeta'(s)}{\zeta(s)}=\sum_{n\ge 1}\frac{\Lambda(n)}{n^s}$.

In the studies of $1/f$ noise, the fast Fourier transform
(FFT)plays a central role. But the FFT refers to the fast
calculation of the discrete Fourier transform (DFT) with a
f\mbox{}inite period $q=2^l$, $l$ a positive integer. In the DFT
one starts with all $q^{\rm{th}}$ roots of the unity $\exp(2i\pi
p/q)$, $p=1\ldots q $ and the signal analysis of the arithmetical
sequence $x(n)$ is performed by projecting onto the $nth$ powers
(or characters of \textit{Z}/q\textit{Z}) with well known
formulas.

The signal analysis based on the DFT is not well suited to
aperiodic sequences with many resonances (naturally a resonance is
a primitive root of the unity: $(p,q)=1$), and the FFT may fail to
discover the underlying structure in the spectrum. We recently
introduced a new method based on the Ramanujan sums defined in
(\ref{eq2}) \cite{Planat4}.

Mangoldt function is related to M\"{o}bius function thanks to the
Ramanujan sums expansion found by Hardy \cite{Gadiyar}
\begin{equation}
b(n)=\frac{\phi(n)}{n}\Lambda(n)=\sum_{q=1}^{\infty}\frac{\mu(q)}{\phi(q)}c_q(n).
\label{equab}
\end{equation}
We call such a type of Fourier expansion a Ramanujan-Fourier
transform (RFT). General formulas are given in our recent
publication \cite{Planat4} and in the paper by Gadiyar
\cite{Gadiyar}. This author also reports on a stimulating
conjecture relating the autocorrelation function of $b(n)$ and the
problem of pairs of prime numbers. In the special case
(\ref{equab}), it is clear that $\mu(q)/\phi(q)$ is the RFT of the
modif\mbox{}ied Mangoldt sequence $b(n)$.
\begin{figure}[htbp]
\centering{\resizebox{10cm}{!}{\includegraphics{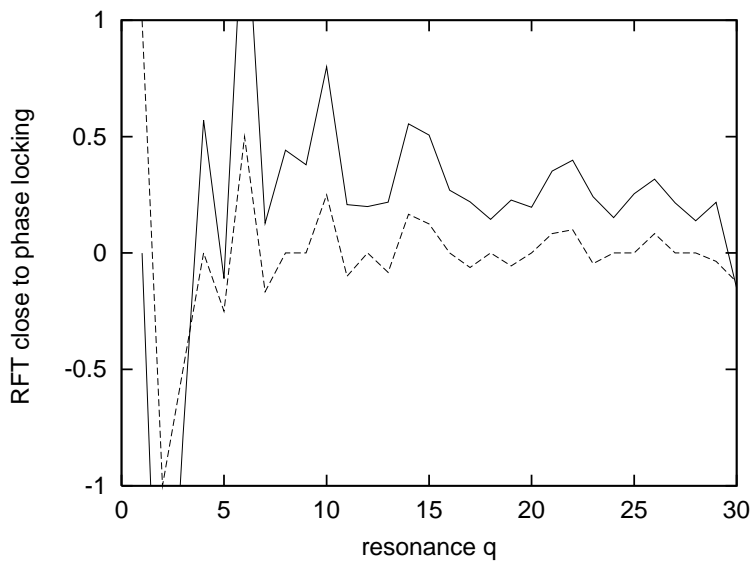}}}
\caption{Ramanujan-Fourier transform (RFT) of the error term
(upper curve) of modif\mbox{}ied Mangoldt function $b(n)$  in
comparison to the function $\mu(q)/\phi(q)$(lower curve). }
\end{figure}
%

%
%
Using Ramanujan-Fourier analysis the $1/f^{2\alpha}$ power
spectrum gets replaced by a new signature shown on Fig. 2, not
very different of $\mu(q)/\phi(q)$ (up to a scaling factor).

\subsection{The hyperbolic geometry of phase noise and $1/f$ frequency noise}

The whole theory can be justified by studying the noise in the
half plane $\it{H}=\{z=\nu+iy,~i^2=-1\footnote{The imaginary
symbol $i$ should not be confused with the index $i$ in integers
$p_i$, $q_i$ and in related integers.},~y>0\}$ of coordinates
$\nu=\frac{f}{f_0}$ and $ y=\frac{f_B}{f_c}>0$ and by introducing
the modular transformations
\begin{equation}
z \rightarrow \gamma(z)=z'=\frac{p_i z+p'_i}{q_i z+q'_i},
~~p_iq'_i-p'_iq_i=1. \label{equation13}
\end{equation}
The set of images of the f\mbox{}iltering line $z=\nu+i$ under all
modular transformations can be written as
\begin{equation}
|z'-(\frac{p_i}{q_i} + \frac{i}{2q_i^2})|=\frac{1}{2q_i^2}.
\label{Ford}
\end{equation}
Equation (\ref{Ford}) def\mbox{}ines Ford circles (see Fig. 3)
centered at points $z=\frac{p_i}{q_i}+\frac{i}{2q_i^2}$ with
radius $\frac{1}{2q_i^2}$ \cite{Rademacher}. To each
$\frac{p_i}{q_i}$ a Ford circle in the upper half plane can be
attached, which is tangent to the real axis at
$\nu=\frac{p_i}{q_i}$. Ford circles never intersect: they are
tangent to each other if and only if they belong to fractions
which are adjacent in the Farey sequence
$\frac{0}{1}<\cdots\frac{p_1}{q_1}<\frac{p_1+p_2}{q_1+q_2}<\frac{q_1}{q_2}\cdots<\frac{1}{1}$\cite{Rademacher}.
\begin{figure}[htbp]
\centering{\resizebox{8cm}{!}{\includegraphics{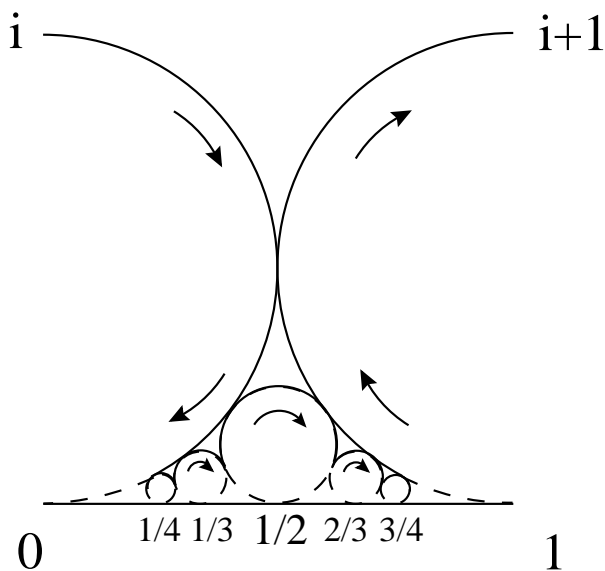}}}
\caption{Ford circles: the mapping of the f\mbox{}iltering line
under modular transformations (\ref{equation13}). The arrows
indicates that Ford circles were used as an integration path by
Rademacher to compute the partition function $p(n)$
\cite{Rademacher}.}
\end{figure}
%
%
%
The half plane $\it{H}$ is the model of Poincar\'{e} hyperbolic
geometry. A basic fact about the modular transformations
(\ref{equation13}) is that they form a discontinuous group
$\Gamma\simeq SL(2,\it{Z})/\{\pm1\}$, which is called the modular
group. The action of $\Gamma$ on the half-plane $\it{H}$ looks
like the one generated by two independent linear translations on
the Euclidean plane, which is equivalent to  a tesselation the
complex plane $\it{C}$ with congruent parallelograms. One
introduces the fundamental domain of $\Gamma$ (or modular surface)
$\it{F}=\{z \in \it{H}:~|z|\ge 1,~|\nu|\le \frac{1}{2}\}$, and the
family of domains $\{\gamma(\it{F}),\gamma \in \Gamma\}$ induces a
tesselation of $\it{H}$ \cite{Gutzwiller}.

It can be shown \cite{Planat3},\cite{Gutzwiller} that the noise
amplitude is a particular type of solution of the eigenvalue
problem with the non-Euclidean Laplacian
$\Delta=y^2(\frac{\partial^2}{\partial
\nu^2}+\frac{\partial^2}{\partial y^2})$. The solution corresponds
to the scattering of waves in the fundamental domain $F$. It can
be approximated as a superposition of three contributions. The
first one is an horizontal wave and is of the power law form
$y^s$, the second one is also horizontal wave of the form $S(s)
y^{1-s}$ and corresponds to a reflected wave with a scattering
coefficient of modulus $|S(s)|=1$, whereas the remaining part
$T(y,\nu)$ is a complex superposition of waves depending of $y$
and the harmonics of $\exp(2i\pi\nu)$, but going to zero for $y
\rightarrow \infty$. Extracting the smooth part in $S(s)$ one is
left with a random factor which is precisely equal to the function
$Z(2s-1)$ defined in (\ref{equation11}) as the quotient of two
Riemann zeta functions at $2s-1$ and $2s$, respectively. An
interesting case is when $s$ is on the critical line, i.e.
$s=\frac{1}{2}+ik$ in which case the superposition $T(y,\nu)$
vanishes and the reflexion coefficient is
\begin{equation}
S(k)= \exp[2i\theta(k)],~~~\mbox{with}~\theta'(k)=\frac{d\ln
S(s)}{ds}~~\mbox{at}~s=\frac{1}{2}+ik.
 \label{hyperbphase}
\end{equation}
The scattering of waves from the modular surface is thus similar
to the phase-locking model plotted in the set of equations
(\ref{equation1})-(\ref{equation11}). It explains the relationship
between the hyperbolic phase and the $1/f$ noise found in the
counting function $\theta'(k)$. The phase factor $\theta(k)$ is
represented in Fig. 4.
\begin{figure}[htbp]
\centering{\resizebox{10cm}{!}{\includegraphics{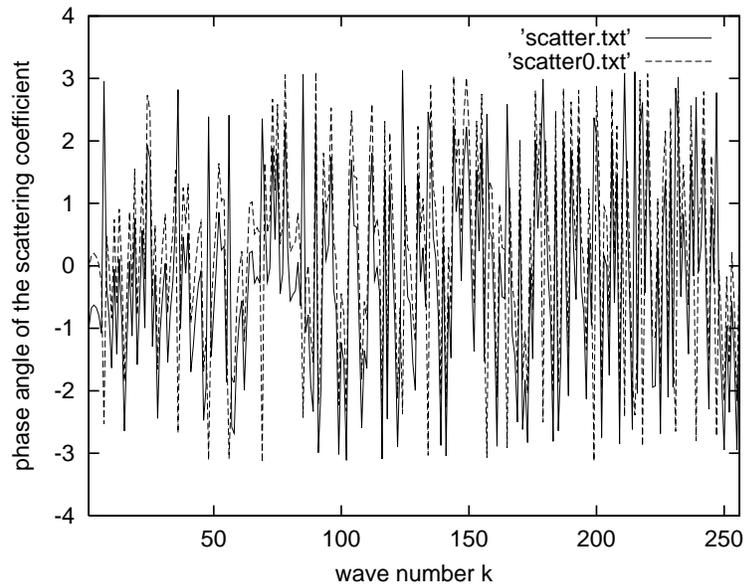}}}
\caption{The phase angle $\theta(k)$ for the scattering of noise
waves on the modular surface. Plain lines: Exact phase factor.
Dotted lines: Approximation based on the quotient of two Riemann
zeta functions\cite{Planat3} .}
\end{figure}
%
%
%

\section{Quantum phase-locking}
\subsection{The quantum phase operators}

Going back to the quantum definition of phase states announced in
the introduction one calculates the projection operator over the
subset of phase-locked quantum states $|\theta' _p\rangle$ as

\begin{equation}
P_q^{\rm{lock}}=\sum_p|\theta'_p\rangle
\langle\theta'_p|=\frac{1}{q}\sum_{n,l} c_q(n-l)|n\rangle \langle
l|,
 \label{eq3}
\end{equation}
where the range of values of $n,l$ is from $0$ to $\phi(q)$. Thus
the matrix elements of the projection are $q\langle n|P_q|l\rangle
= c_q(n-l)$. This sheds light on the equivalence between
cyclotomic lattices of algebraic number theory and the quantum
theory of phase-locked states.

The projection operator over the subset of pairs of phase-locked
quantum states $|\theta' _p\rangle$ is calculated as
\begin{equation}
P_q^{\rm{pairs}}=\sum_{p,\bar{p}}|\theta'_p\rangle
\langle\theta'_{\bar{p}}|=\frac{1}{q}\sum_{n,l} k_q(n,l)|n\rangle
\langle l|,
 \label{eq3bis}
\end{equation}
where the notation $p,\bar{p}$ means that the summation is applied
to such pairs of states satisfying (\ref{phaselocked}). The matrix
elements of the projection are $q\langle
n|P_q^{\rm{pairs}}|l\rangle = k_q(n,l)$, which are in the form of
so-called Kloosterman sums \cite{Terras99}
\begin{equation}
k_q(n,l)=\sum_{p,\bar{p}}\exp[\frac{2i\pi}{q}(pn-\bar{p}l)].
\label{Kloos}
\end{equation}
Kloosterman sums $k_q(n,l)$ as well as Ramanujan sums $c_q(n-l)$
are relative integers. They are given below for the Hilbert
dimensions $q=5 (\phi(5)=4)$ and $q=6(\phi(6)=2)$.
\begin{eqnarray}
&q=5:~~ c_5=\left[
\begin{array}{cccc}
~~4 & -1&-1&-1 \\
-1 & ~~4&-1&-1\\
-1&-1&~~4&-1\\
-1&-1&-1&~~4
\end{array}
\right],~~~~k_5=\left[
\begin{array}{cccc}
-1 & -1&-1&~~4 \\
-1 & ~~4&-1&-1\\
-1&-1&~~4&-1\\
~~4&-1&-1&-1
\end{array}
\right],\nonumber\\
&q=6:~~c_6=\left[
\begin{array}{cc}
 2&1\\
1&2
 \end{array}
 \right],~~k_6=\left[
 \begin{array}{cc}
 -1&~~2\\
~~-2&1
\end{array}
\right].\nonumber
\end{eqnarray}
One de$\rm{fi}$nes the quantum phase-locking operator as
\begin{equation}
\Theta_q^{\rm{lock}}=\sum_p \theta_p |\theta'_p\rangle
\langle\theta'_p|=\pi
P_q^{\rm{lock}}~~\rm{with}~\theta_p=2\pi\frac{p}{q}. \label{eq4}
\end{equation}
The Pegg and Barnett operator \cite{Pegg} is obtained by removing
the coprimality condition. It is Hermitian with eigenvalues
$\theta_p$. Using the number operator $N_q=\sum_{n=0}^{q-1} n|n
\rangle \langle n|$ a generalization of Dirac's commutator
$[\Theta_q,N_q]=-i$ has been obtained.

 Similarly one de$\rm{fi}$nes the quantum phase operator for Kloosterman pairs as
\begin{equation}
\Theta_q^{\rm{pairs}}=\sum_{p,\bar{p}} \theta_p |\theta'_p\rangle
\langle\theta'_{\bar{p}}|=\pi
P_q^{\rm{pairs}}~~\rm{with}~\theta_p=2\pi\frac{p}{q}.
\label{eq4bis}
\end{equation}

The phase number commutator for phase-locked states calculated
from (\ref{eq4}) is
\begin{equation}
C_q^{\rm{lock}}=[\Theta_q^{\rm{lock}},N_q]=\frac{\pi}{q}\sum_{n,l}(l-n)c_q(n-l)|n\rangle\langle
l|, \label{eq5}
\end{equation}
with antisymmetric matrix elements $\langle
l|C_q^{\rm{lock}}|n\rangle=\frac{\pi}{q}(l-n)c_q(n-l)$.

For pairs of phase-locked states an antisymmetric commutator
$C_q^{\rm{pairs}}$ similar to (\ref{eq5}) is obtained with
$k_q(n,l)$ in place of $c_q(n-l)$.

\subsection{Phase expectation value and variance}

 The $\rm{fi}$nite quantum mechanical rules are encoded in the
expectation values of the phase operator and phase variance.

\begin{figure}[htbp]
\centering{\resizebox{10cm}{!}{\includegraphics{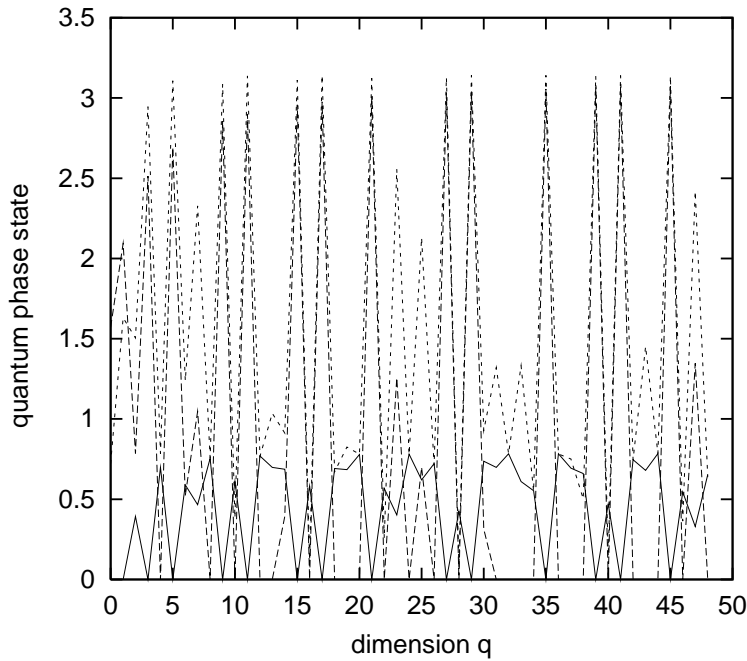}}}
\caption{Oscillations in the expectation value (\ref{expec2}) of
the locked phase at $\beta=1$ (dotted line) and their squeezing at
$\beta=0$ (plain line). The brokenhearted line which touches the
horizontal axis is $\pi \Lambda(q)/\ln q$.}
\end{figure}
%
%
%
%
\begin{figure}[htbp]
\centering{\resizebox{10cm}{!}{\includegraphics{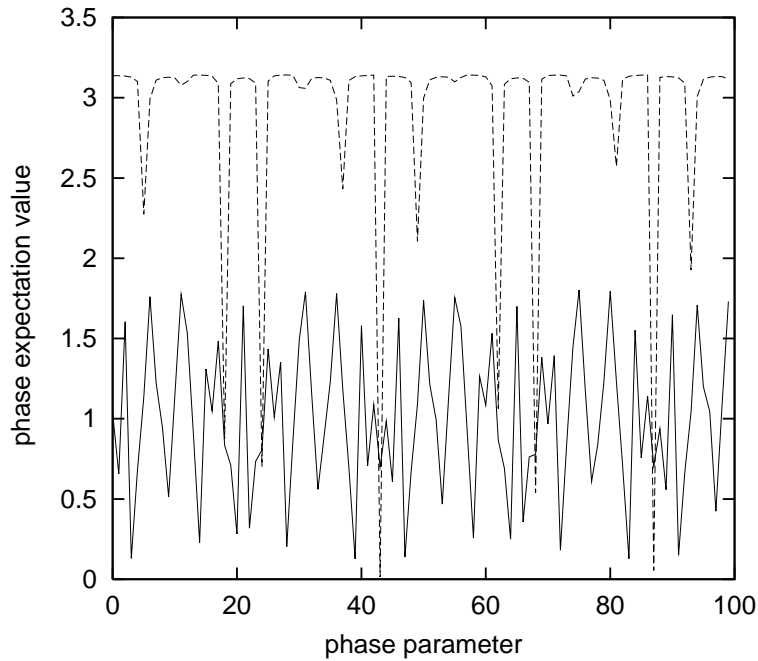}}}
\caption{Phase expectation value versus the phase parameter
$\beta$. Plain lines: $q=15$. Dotted lines $q=13$.}
\end{figure}
%

%
Rephrasing Pegg and Barnett, let us consider a pure phase state
 $|f \rangle = \sum_{n=0}^{q-1} u_n |n \rangle$ having $u_n$ of the form
\begin{equation}
u_n=(1/\sqrt{q})\exp(i n\beta), \label{pure}
\end{equation}
where $\beta$ is a real phase parameter. One de$\rm{fi}$nes the
phase probability distribution $\langle\theta'_p|f\rangle^2$, the
phase expectation value $\langle
\Theta_q^{\rm{lock}}\rangle=\sum_{p}\theta_p
\langle\theta'_p|f\rangle^2$, and the phase variance $(\Delta
\Theta_q ^2)^{\rm{lock}}=\sum_p (\theta_p-\langle
\Theta_q^{\rm{lock}}\rangle)^2 \langle\theta'_p|f\rangle^2$. One
gets
\begin{eqnarray}
&\langle\Theta_q^{\rm{lock}}\rangle=\frac{\pi}{q^2}\sum_{n,l}
c_q(l-n) \exp[i\beta(n-l)],\label{expec2}\\
&(\Delta \Theta_q^2)^{\rm{lock}}=4 \langle
\tilde{\Theta}_q^{\rm{lock}}\rangle
+\frac{\langle\Theta_q\rangle^2}{\pi}(\langle\Theta_q\rangle-2\pi),
\end{eqnarray}
with the modi$\rm{fi}$ed expectation value
$\langle\tilde{\Theta}_q^{\rm{lock}}\rangle=\frac{\pi}{q^2}\sum_{n,l}
\tilde{c}_q(l-n) \exp[i\beta(n-l)]$,
 and the modi$\rm{fi}$ed Ramanujan sums
 $\tilde{c}_q(n)=\sum_p (p/q)^2 \exp(2i\pi
m\frac{p}{q})$.

Fig. 5 illustrates the phase expectation value versus the
dimension $q$ for two different values of the phase parameter
$\beta$. For $\beta=1$ they are peaks at dimensions $q=p^r$ which
are powers of a prime number $p$. The most significant peaks are
fitted by the function $\pi\Lambda(q)/\ln q$, where $\Lambda(q)$
is the Mangoldt function introduced in (\ref{Riemann}) of Sect.2.
This observation provides the link between the arithmetical
hyperbolic viewpoint and the quantum one. A deepest explanation
based on the relation with quantum statistical mechanics and the
work of Bost and Connes can be found in \cite{PlanatKMS}. For
$\beta=0$ the peaks are smoothed out due to the averaging over the
Ramanujan sums matrix. Fig. 6 shows the phase expectation value
versus the phase parameter $\beta$. For the case of the prime
number $q=13$, the mean value is high with absorption like lines
at isolated values of $\beta$. For the case of the dimension
$q=15$ which is not a prime power the phase expectation is much
lower in value and much more random.

Fig. 7 illustrates the phase variance versus the dimension $q$.
Again the case $\beta=1$ leads to peaks at prime powers. Like the
expectation value in Fig. 5, it is thus reminiscent of the
Mangoldt function. Mangoldt function $\Lambda(n)$ is
de$\rm{fi}$ned as $\ln p$ if n is the power of a prime number $p$
and $0$ otherwise. It arises in the frame of prime number theory
\cite{Planat1} from the logarithmic derivative of the Riemann zeta
function $\zeta(s)$ as
$-\frac{\zeta'(s)}{\zeta(s)}=\sum_{n=0}^{\infty}\frac{\Lambda(n)}{n^s}$.
Its average value oscillates about $1$ with an error term which is
explicitely related to the positions of zeros of $\zeta(s)$ on the
critical line $s=\frac{1}{2}$. The error term shows a power
spectral density close to that of $1/f$ noise \cite{Planat1}. It
is stimulating to recover results reminding prime number theory in
the new context of quantum phase-locking.

 Finally, the phase variance is considerably smoothed
out for $\beta=\pi$ and is much lower than the classical limit
$\pi^2/3$. The parameter $\beta$ can thus be interpreted as a
squeezing parameter since it allows to de$\rm{fi}$ne quantum
phase-locked states having weak phase variance for a whole range
of dimensions.
\begin{figure}[htbp]
\centering{\resizebox{10cm}{!}{\includegraphics{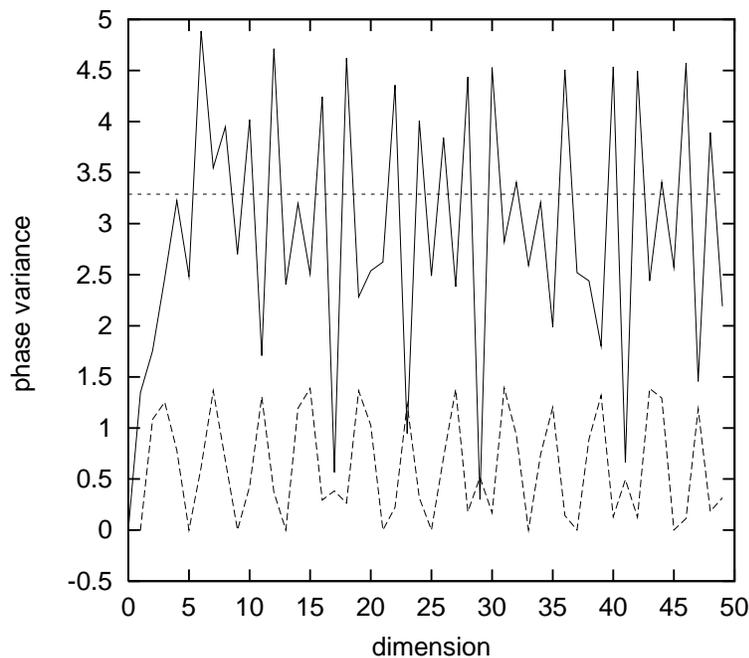}}}
\caption{Phase variance versus the dimension q of the Hilbert
space. Plain lines: $\beta=1$. Dotted lines: $\beta=\pi$.}
\end{figure}
\begin{figure}[htbp]
\centering{\resizebox{10cm}{!}{\includegraphics{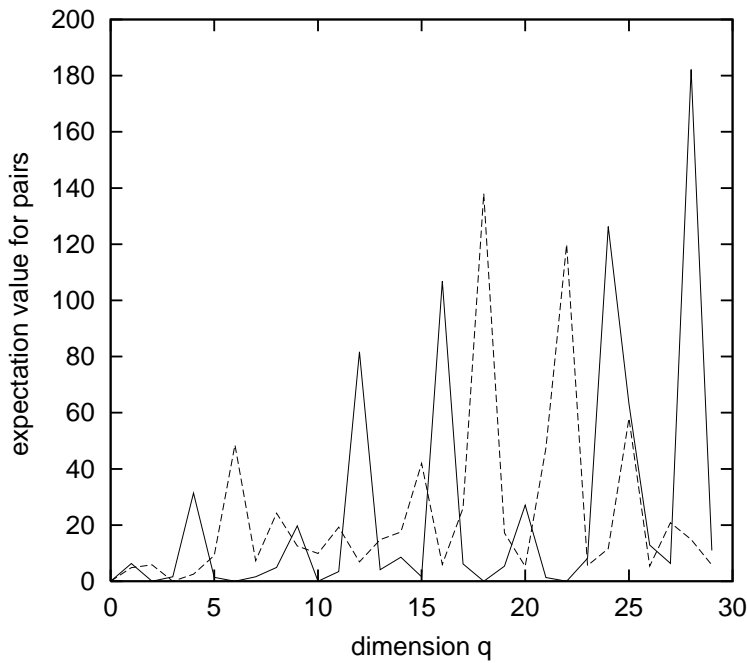}}}
\caption{Phase expectation value versus the dimension $q$ for
pairs of phase-locked states. Plain lines: $\beta=0$. Dotted
lines: $\beta=1$.}
\end{figure}
%
%
%
%

\subsection{Towards discrete phase entanglement} The expectation value of
quantum phase states can be rewritten using the projection
operator of individual phase states $\pi_p=|\theta'_p\rangle
\langle \theta'_p|$ as follows
\begin{equation}
\langle \Theta_q^{\rm{lock}}\rangle=\sum_p \theta_p \langle
f|\theta'_p \rangle \langle \theta'_p |f \rangle=\sum_p \theta_p
\langle f|\pi_p|f \rangle.
\end{equation}
This suggests a de$\rm{fi}$nition of expectation values for pairs
based on the product $\pi_p \pi_{\bar{p}}$ as follows
\begin{equation}
\langle \Theta_q^{\rm{pairs}}\rangle=\sum_{p,\bar{p}} \theta_p
\langle f|\pi_p \pi_{\bar{p}}|f \rangle . \label{pairs}
\end{equation}
It is inspired by the quantum calculation of correlations in
Bell's theorem \cite{Scully}. Using pure phase states as in
(\ref{pure}) we get
\begin{equation}
\langle \Theta_q^{\rm{pairs}}\rangle=\frac{2
\pi}{q^2}\sum_{n,l}\tilde{k_q}(n,l) \exp[i \beta(n-l)],
\end{equation}
where we introduced generalized Kloosterman sums\\
$\tilde{k}_q(n,l)=\sum_{p,\bar{p}} p \exp[\frac{2 i
\pi}{q}(p-\bar{p})(l-n)]$. These sums are in general complex
numbers (and are not Gaussian integers). The expectation value is
real as expected. In Fig. 8 it is represented versus the dimension
$q$ for two different values, $\beta=0$ and $\beta=1$,
respectively. Note that the pair correlation (\ref{pairs}) is very
strongly dependent on $q$ and becomes quite huge at some values.

This result suggests that a detailed study of Bell's type
inequalities based on quantum phase-locked states, and their
relationship to the properties of numbers, should be undertaken.
Calculations involving fully entangled states
\begin{equation}
|f\rangle= \frac{1}{q}\sum_{p,\bar{p}} |\theta_p,1\rangle \otimes
|\theta_{\bar{p}},2\rangle,
\end{equation}
have to be carried out. This is left for future work.

\subsection{The discrete phase: cycles in $Z/qZ$} There is a scalar viewpoint for the
above approach, which emphasizes well the intricate order of the
group $Z/qZ$, the group of integers modulo $q$. One asks the
question: what is the largest cycle in that group. For that
purpose one looks at the primitive roots, which are the solutions
$g$ of the equation
\begin{equation}
g^{\alpha} \equiv 1(\textrm{mod}~q),
\end{equation}
such that the equation is wrong for any $1 \le \alpha < q-1$ and
true only for $\alpha=q-1$. If q=p, a prime number, and $p=7$, the
largest period is thus $\phi(p)$=p-1=6, and the cycle is as given
in Table I. If $q=2$, $4$, $q=p^r$, a power a prime number $>2$,
or $q=2p^r$, twice the power of a prime number $>2$, then a
primitive root exists, and the largest cycle in the group is
$\phi(q)$. For example $g=2$ and $q=3^2$ leads to the period
$\phi(9)=6<q-1=8$, as shown in Table II.
\begin{table}[t]
\caption{$(Z/7Z)^*$ is a cyclic group of order $\phi(7)=6$.}
\begin{center} \footnotesize
\begin{tabular}{|c|c|c|c|c|c|c|c|c|}
\hline {$\alpha$}     & {1} &{2} &{3} &{4} &{5} &{6} &{7} &{8}  \\
\hline {$3^{\alpha}$} & {3} &{2} &{6} &{4} &{5} &{1} &{3} &{2}\\
\hline
\end{tabular}
\end{center}
\end{table}
\begin{table}[t]
\caption{$(Z/3^2 Z)^*$,  is a cyclic group of order $\phi(9)=6$.}
\begin{center}
\footnotesize
\begin{tabular}{|c|c|c|c|c|c|c|c|c|}
\hline {$\alpha$}     & {1} &{2} &{3} &{4} &{5} &{6} &{7} &{8}  \\
\hline {$2^{\alpha}$} & {2} &{4} &{8} &{7} &{5} &{1} &{2} &{4}\\
\hline
\end{tabular}
\end{center}
\end{table}
Otherwise there is no primitive root. The period of the largest
cycle in $Z/qZ$ can still be calculated and is called the
Carmichael Lambda function $\lambda(q)$. It is shown in Table III
for the case $g=3$ and $q=8$. It is
$\lambda(8)=2<\phi(8)=4<8-1=7$.
\begin{table}[t]
\caption{$(Z/8Z)^*$ has a largest cyclic group of order
$\lambda(8)=2$.}
\begin{center}
\footnotesize
\begin{tabular}{|c|c|c|c|c|c|c|c|c|}
\hline {$\alpha$}     & {1} &{2} &{3} &{4} &{5} &{6} &{7} &{8}  \\
\hline {$2^{\alpha}$} & {3} &{1} &{3} &{1} &{3} &{1} &{3} &{1}\\
\hline
\end{tabular}
\end{center}
\end{table}
\begin{figure}[htbp]
\centering{\resizebox{8cm}{!}{\includegraphics{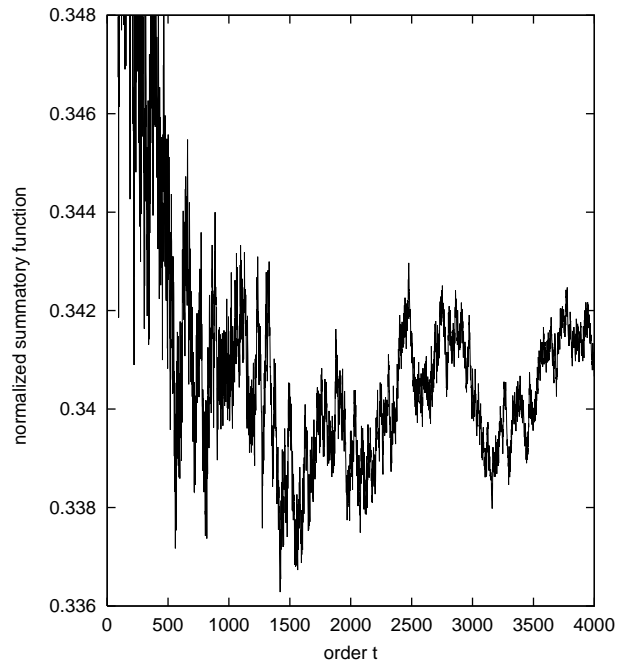}}}
\caption{Normalized Carmichael lambda function: $(\sum_1^t
\lambda(n))/t^{1.90}$.}
\end{figure}
\begin{figure}[htbp]
\centering{\resizebox{8cm}{!}{\includegraphics{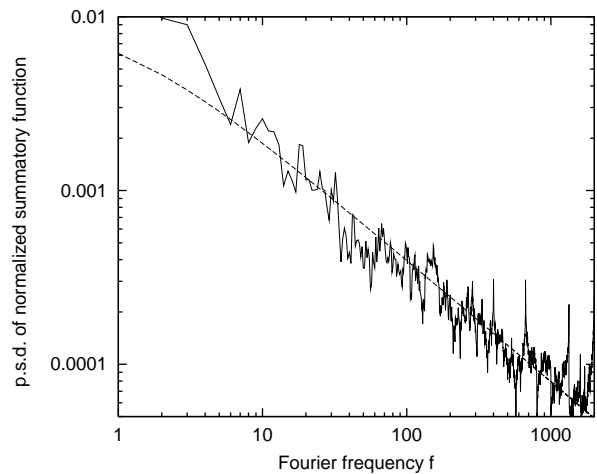}}}
\caption{FFT of the normalized Carmichael lambda function. The
staight line has slope $-0.70$.}
\end{figure}
%
%
%
Fig. 9 shows the properly normalized period for the cycles in
Z/qZ. Its fractal character can be appreciated by looking at the
corresponding power spectral density shown in Fig. 10. It has the
form of a $1/f^{\alpha}$ noise, with $\alpha=0.70$. For a more
refined link between primitive roots $g$, cyclotomy and Ramanujan
sums see also \cite{Moree}.

\section{Conclusion} In conclusion, we explained how useful could
be the concepts of prime number theory in explaining various
features of phase-locking at the classical and quantum level. In
the classical realm we reminded the hyperbolic geometry of phase,
which occurs when one accounts for all harmonics in the mixing and
low-pass filtering process, how $1/f$ frequency noise is produced
and how it is related to Mangoldt function, and thus to the
critical zeros of Riemann zeta function. Then we studied several
properties resulting from introducing phase-locking in
Pegg-Barnett quantum phase formalism. The idea of quantum
teleportation was initially formulated by Bennett et al in
$\rm{fi}$nite-dimensional Hilbert space \cite{Bennett}, but, yet
independently of this, one can conjecture that cyclotomic aspects
in phase-locking could play an important role in many fundamental
tests of quantum mechanics related to quantum entanglement. Munro
and Milburn \cite{Munro} already conjectured that the best way to
see the quantum nature of correlations in entangled states is
through the measurement of the observable canonically conjugate to
photon number, i.e. the quantum phase. In their paper dealing with
the Greenberger-Horne-Zeilinger quantum correlations, they
presented a homodyne scheme requiring discrete phase measurement.
We expect that the interplay between quantum mechanics and number
theory will appear repetitively in the coming attempts to
manipulate quantum information \cite{Wootters}.

\section*{Acknowledgments}
The third part of this paper was presented at the International
Conference on Squeezed States and Uncertainty Relations in Puebla,
in June 2003. The authors acknowledge Hector Moya for his
invitation.

\end{document}